\documentclass[journal]{IEEEtran}
\ifCLASSINFOpdf
\else
\fi
\usepackage{amsmath}
\DeclareMathOperator*{\argmin}{argmin} 
\usepackage{algorithm}
\usepackage[noend]{algpseudocode}
\usepackage{makecell}
\usepackage{amssymb,graphicx}
\usepackage{tikz}
\usepackage[centerlast,small,sc]{caption}
\usetikzlibrary{positioning}
\usetikzlibrary{decorations.pathreplacing,calligraphy}

\newcommand{\myfW}{\ensuremath{\boldsymbol{f}_{\mathbf{W}}}}
\newcommand{\myg}{\ensuremath{\boldsymbol{g}}}

\newcommand{\myM}{\ensuremath{\mathbf{M}}}
\newcommand{\mynabla}{\ensuremath{\boldsymbol{\nabla}}}

\newcommand{\myPhi}{\ensuremath{\boldsymbol{\Phi}}}
\newcommand{\myS}{\ensuremath{\mathbf{S}}}

\newcommand{\mytheta}{\ensuremath{\boldsymbol{\theta}}}
\newcommand{\myu}{\ensuremath{\boldsymbol{u}}}
\newcommand{\myW}{\ensuremath{\mathbf{W}}}
\newcommand{\myx}{\ensuremath{\boldsymbol{x}}}

\begin{document}
%
\title{Learning to Solve Inverse Problems \\ for Perceptual Sound Matching}
%
%
%

\author{Han~Han,~\IEEEmembership{Student Member,~IEEE,}   Vincent~Lostanlen,~\IEEEmembership{Member,~IEEE}      and~Mathieu~Lagrange,~\IEEEmembership{Member,~IEEE}

\thanks{H. Han, V. Lostanlen, and M. Lagrange are with Nantes Université, École Centrale Nantes, CNRS, LS2N, UMR 6004, F-44000 Nantes, France. e-mail: (han.han; vincent.lostanlen; mathieu.lagrange@ls2n.fr).}
}

%
%

\markboth{Transactions on Audio, Speech, and Language Processing,~Vol.~X, No.~Y, submitted~October~2023, revised~April~2024}%
{Han \MakeLowercase{\textit{et al.}}: \title{}}
%



\maketitle

\begin{abstract}
Perceptual sound matching (PSM) aims to find the input parameters to a synthesizer so as to best imitate an audio target.
Deep learning for PSM optimizes a neural network to analyze and reconstruct prerecorded samples.
In this context, our article addresses the problem of designing a suitable loss function when the training set is generated by a differentiable synthesizer.
Our main contribution is perceptual--neural--physical loss (PNP), which aims at addressing a tradeoff between perceptual relevance and computational efficiency.
The key idea behind PNP is to linearize the effect of synthesis parameters upon auditory features in the vicinity of each training sample.
The linearization procedure is massively parallelizable, can be precomputed,  and offers a 100-fold speedup during gradient descent compared to differentiable digital signal processing (DDSP).
We demonstrate PNP on two datasets of nonstationary sounds: an AM/FM arpeggiator and a physical model of rectangular membranes.
We show that PNP is able to accelerate DDSP with joint time--frequency scattering transform (JTFS) as auditory feature while preserving its perceptual fidelity. 
Additionally, we evaluate the impact of other design choices in PSM: parameter rescaling, pretraining, auditory representation, and gradient clipping.
We report state-of-the-art results on both datasets and find that PNP-accelerated JTFS has greater influence on PSM performance than any other design choice.

\end{abstract}


%
\IEEEpeerreviewmaketitle

\section{Introduction}
\IEEEPARstart{I}{mitation} plays a central role in intelligent sonic interactions. 
In particular, the sensorimotor skills underlying vocal learning are acquired by repeated imitation of tutors. 
Another example is given by music: by aiming at an exact replication of a teacher's performance, students gradually refine their level of technical control on the instrument. 
In both scenarios, two cognitive abilities form a closed loop: listening and search. 
On one hand, listening evaluates the perceptual fidelity of imitation by comparing each sonic replica with its reference.
On the other hand, search gradually refines the execution of the replica at the gestural level.

\subsection{Motivation}
Endowing machines with the general-purpose ability to imitate sounds by fine motor control, as opposed to simple loudspeaker playback, has the potential to transform research protocols in robotics and human--computer interaction \cite{oudeyer:hal-00818204}. 
In this context, we call \emph{perceptual sound matching} (PSM) the task of retrieving the input parameters of a synthesizer so as to best imitate a digital audio target.
PSM is an emerging subdomain of artificial intelligence (AI) whose application scope is rapidly expanding, thanks to the recent development of highly expressive synthesizers for speech \cite{wangspeech2020,liu20_interspeech}, music \cite{renault:hal-04073770,WigginsK23}, and bioacoustics \cite{amador2021}. 

Procedural audio synthesis is found at multiple levels of complexity: from digital audio synthesizer parameters,  tangible parameters reflecting physical constructions and gestures, to polyphonic music notations for multi-player ensemble.  
Consider a target sound producible by a commercial synthesizer in music production and sound design: deriving the optimal set of input parameters that reproduce this sound accelerates and eases the synthesizer control \cite{caspe2022ddx7, Barkan2019}.
When the synthesizer is a physical model \cite{traube2001,riionheimo2003}, inferring the physical parameters is an interesting inverse problem that reveals the excitation mechanism as well as physical constructs of the sounding bodies, paving ways for digital musical instrument design (DMI). 
Even more challenging is the case of automatic orchestration \cite{cella2020}, where parameters carry physical meaning of instrument type, playing techniques, dynamics etc. 

\subsection{Problem statement}

Let us denote a target sound by the signal $\myx$, a parametric synthesizer by the function $\myg$, and its search space by the set $\Theta$.
We define PSM in terms of a function $\boldsymbol{f}$ which takes $\myx$ as input and returns an element of $\Theta$, denoted by $\tilde{\mytheta} = \boldsymbol{f}(\boldsymbol{x})$.
For an ideal PSM, the replica $\tilde{\myx} = \myg(\tilde{\mytheta}) = (\myg \circ \boldsymbol{f})(\myx)$ is indiscernible from the target sound $\myx$.
On the contrary, failure cases of PSM arise when $\myx$ and $\tilde{\myx}$ differ beyond the psychoacoustic threshold of just noticeable differences: the minimal change needed for a difference to be perceived.
Thus, the computational evaluation of PSM requires defining a psychoacoustic descriptor $\myPhi$ such that $\myPhi(\myx) = \myPhi(\tilde{\myx})$ if and only if $\myx$ and $\tilde{\myx}$ are indiscernible.
Throughout this paper, we assume the function $(\myPhi \circ \myg)$ to be differentiable and deterministic.
Given $\myPhi$, $\myg$, and $\Theta$, the optimal PSM $\boldsymbol{f}$ appears as the solution of the following minimization problem:
 \begin{equation}
     \boldsymbol{f}(\myx) \in \underset{\tilde{\mytheta} \in \Theta}\argmin\; \big\Vert (\myPhi \circ \myg)(\tilde{\mytheta})-\myPhi(\myx) \big\Vert^2_2.
     \label{eq:psm}
\end{equation}


Given $\myx$, if there exists some $\mytheta$ such that $\myx = \myg(\mytheta)$, then setting $\tilde{\mytheta} = \mytheta$ in the equation above leads to $\myg(\tilde{\mytheta})=\myx$ and thus $\Vert (\myPhi \circ \myg)(\tilde{\mytheta})-\myPhi(\myx) \Vert^2_2 = 0$.
Thus, $\mytheta$ is an optimal PSM value for $\boldsymbol{f}(\myx)$.
Yet, in full generality, no such $\mytheta$ exists and the optimum of Equation \ref{eq:psm} is reached at a nonzero perceptual distance: the associated inverse problem is said to be \emph{ill-posed}.

The search space $\Theta$ of a synthesizer is theoretically infinite; in practice, it may contain up to $10^9$ elements \cite{turian2021one}.
Furthermore, the computation of a multidimensional descriptor $\myPhi$ that is suitable to represent timbre dissimilarities is costly, even on GPU hardware.
For these two reasons, retrieving the optimal PSM value $\boldsymbol{f}(\myx)$ by exhaustive evaluation of $(\myPhi\circ\myg)$ over $\Theta$ is intractable for any problem of realistic size.
Instead, prior publications have resorted to iterative methods such as genetic algorithms \cite{yeeking2011}
or gradient descent \cite{vahidi2023meso} so as to update $\tilde{\mytheta}$ from a random guess until it reaches convergence.
Though reliable solutions may still be obtained this way, systems relying on iterative methods require step-wise reevaluation of the perceptual descriptor and new optimization steps each time given a different sound query. Moreover, given a nonconvex objective, there is no guarantee of converging to the same optimal solution across queries.

On the contrary, training a deep neural network (DNN) for PSM circumvents the aforementioned problems of uncertain and costly predictions. 
Denoting its weights by $\myW$, the prediction of $\tilde{\mytheta}$ from $\myx$ incurs a single forward pass: i.e, $\tilde{\mytheta}=\myfW(\myx)$: the computational cost of this prediction depend solely on the DNN architecture and not on $\myPhi$ nor $\myg$.
Yet, training $\myfW$ to minimize $\Vert (\myPhi \circ \myg \circ \myfW) (\myx)-\myPhi (\myx) \Vert_2^2$ remains constrained by the cost of $(\myPhi \circ \myg )$.
Our article offers a generic solution to mitigate that constraint.

\subsection{Key idea}

\begin{figure}
\begin{tikzpicture}[thick]
\node(theta){\mytheta};
\node(physical)[above=of theta, align=left, yshift=-9mm]{\scriptsize Parametric\\\scriptsize domain};
\node(x)[label={[anchor=west, xshift=-2mm, yshift=-5mm]\scriptsize  original}, right=of theta]{\myx};
\node(audio)[above=of x, align=left, yshift=-9mm]{\scriptsize Audio\\\scriptsize domain};
\node(S)[right=of x]{\myS};
\node(perceptual)[above=of S, align=left, yshift=-9mm]{\scriptsize Perceptual\\\scriptsize domain};

\node(thetatilde)[below=of theta]{$\tilde{\mytheta}$};
\node(xtilde)[label={[anchor=west, xshift=-2mm, yshift=-6mm]\scriptsize reconstruction}, right=of thetatilde]{$\tilde{\myx}$};
\node(Stilde)[below=of S]{$\tilde{\myS}$};

\draw[decorate, decoration = {calligraphic brace, mirror}] (3.1,-1.7) --  (3.1,0.25) node(DDSP)[xshift=7mm, pos=0.5, align=left] {\scriptsize  DDSP\\\scriptsize spectral$\;\approx$\\\scriptsize loss};

\draw[decorate, decoration = {calligraphic brace}] (5.5,-1.7) --  (5.5,0.25) node(PNP)[xshift=-6mm, pos=0.5, align=left]{\scriptsize PNP\\\scriptsize quadratic\\\scriptsize form};

\node(thetabis)[xshift=24mm, right=of S]{\mytheta};
\node(xbis)[right=of thetabis]{\myx};
\node(thetatildebis)[below=of thetabis]{$\tilde{\mytheta}$};

\node(M)[right=of DDSP,xshift=4mm]{$\myM(\mytheta)$};
\node(manifold)[above=of M, align=left, yshift=-2mm]{\scriptsize Riemannian\\\scriptsize metric};

\draw[->] (theta) -- (x) node[above,midway]{\myg};
\draw[->] (x) -- (S) node[above,midway]{\myPhi};
\draw[->, dashed] (x) -- (thetatilde) node[above,xshift=-2mm,midway]{\myfW};
\draw[->, dashed] (thetatilde) -- (xtilde) node[above,midway]{\myg};
\draw[->, dashed] (xtilde) -- (Stilde) node[above,midway]{\myPhi};

\draw[->] (thetabis) -- (xbis) node[above,midway]{\myg};
\draw[->, dashed] (xbis) -- (thetatildebis) node[above,xshift=-2mm,midway]{\myfW};
\draw[->] (thetabis) -- (M) node[above,xshift=-5mm,yshift=-1mm,midway]{$\mynabla_{(\myPhi\circ\myg)}$};


\end{tikzpicture}

\caption{Graphical outline of the proposed method.
Given a known synthesizer $\myg$ and feature map $\myPhi$, we train a neural network $\myfW$ to estimate $\tilde{\mytheta}$ and minimize the ``perceptual--neural--physical'' (PNP) quadratic form
$\langle\tilde{\mytheta} - \mytheta\big\vert\myM(\mytheta)\vert\tilde{\mytheta} - \mytheta\rangle$ where $\myM$ is the Riemannian metric associated to $(\myPhi \circ \myg)$.
Transformations in solid (resp. dashed) lines can (resp. cannot) be cached during training.
\label{fig:overview}}
\end{figure}

We propose a novel loss objective termed "perceptual-neural-physical" (PNP) loss. PNP allows working with computationally heavy $\myPhi$ that would otherwise induce intractable training time, under the conditions where $\myPhi$ and $\myg$ are differentiable. It is also required that $\mytheta$ be known to supervise the training with PNP loss.

PNP is the optimal quadratic approximation of a given $L^2$ perceptual loss.
It adopts a bilinear form whose kernel matrix is the Riemannian metric formed by the differentiable map $(\myPhi \circ \myg)$.
These kernel matrices may be precomputed and cached in memory before training. 
Furthermore, we propose a corrective term added onto the kernel matrix, as a mechanism to reduce the condition number of the kernel matrix in case where the inverse problem is ill-conditioned.

\subsection{Contributions}
We train models against various state-of-the-art loss functions to perform sound matching tasks on two differentiable synthesizers - an AM/FM arpeggiator and a rectangular drum physical model. 
The former is a pure tone chirp signal with amplitude-modulated temporal envelope, an extension to our previous work \cite{han2023pnp}. The latter is an inharmonic transient signal with uneven exponential decay envelope on each partial, which better reflects sounds in the real world.

We demonstrate experimentally that optimizing the DNN against our proposed PNP loss based on JTFS consistently achieves the best performance on both sound matching tasks. Meanwhile we show that the tasks are nontrivial given the poor performance with previous state of the art method using MSS spectral loss. 
Finally, we extend \cite{han2023pnp} by presenting the learning dynamics and an ablation study of our proposed model to discuss the key factors contributing to empirical success (or failure).
We supplement this work with open-source code. \footnote{Companion website: https://github.com/lylyhan/perceptual\_neural\_physical}

\section{Related Work}

\subsection{Differentiable Synthesizers}

Synthesizers have longstanding tradition in digital music production. 
They range in design principles but mostly fall into a few families - subtractive, additive, frequency modulation, and wavetable synthesis. 
Using filters to amplify or diminish certain frequency components, using oscillators to modulate another oscillator's pitch, phase, and amplitude envelope, digital synthesizers elicit a rich variety of timbres. 
Powerful as they are, the concept of differentiable synthesizers only came in the advent of differentiable digital signal processing (DDSP) model \cite{engel2020ddsp}.
DDSP models are encoder-decoder architectures capable of generating high-fidelity audio by passing learned encodings through expressive synthesizers.  
Optimizing against a perceptual loss in the form of Equation. \ref{eq:psm}, DDSP model's parameter updates via back propagation depend on computing $\mynabla_{(\myPhi \circ \myg)} (\mytheta)$. 
This computation is made easy by automatic differentiation (AD), the automatic evaluation of compounds of partial derivatives via the chain rule, during forward pass through the function $(\myPhi \circ \myg)$. 
Consequently, in order to be compatible with AD, the forward function composes differentiable functions. 

To enable learning with perception-based objectives, the recent years have seen a surge of newly implemented differentiable synthesizers.
 Such an implementation is straightforward for certain synthesizers, ones that can be expressed as a cascade of linear or nonlinear continuous functions. Additive synthesizers \cite{masuda2023}, subtractive synthesizers \cite{sawsing}, harmonic plus noise synthesis \cite{engel2020ddsp}, FIR and IIR filters with closed form impulse responses, all fall under this category. 

Other synthesizers with discrete elements, thresholding or nonlinear analog effects may be adapted into differentiable implementations as well.
Facing the infinite discrete choices of routing and patching in FM synthesis, 
\cite{caspe2022ddx7} implemented a differentiable six-oscillator FM synthesizer with fixed patching for each target musical instrument, in order to prescribe harmonic spectra and ease the retrieval of continuous control parameters to regenerate instrument sounds. 
Time-varying nonlinear effects particular to analog audio devices that lack an analytic expression have also been modeled by learned neural net, for example phasor effects in \cite{carson2023differentiable}.  
Other interests lie in enabling the differentiability of synthesis based on physical models, due to its physical interpretability and proximity to realistic sounds. Commonly involving solving a partial differential equation, such synthesis algorithms grow increasingly sophisticated and time-consuming as the underlying physical model increases in dimensionality of solution space and complexity of initial conditions.   \cite{parker2022physical} and \cite{diaz2022rigidbody} trained respectively recurrent and convolution-based neural network to model time-stepping solution from finite difference method and resonator responses from finite element method. 
We refer to \cite{hayes2023review} for a comprehensive review of differentiable synthesizers.

Whether through direct conversion or data-driven simulation via learned filters, enabling the differentiability of synthesis algorithms opens up opportunities for optimizations informed by audio samples. 
Nevertheless it is important to note that, in face of sophisticated synthesis algorithms $\myg$ and cumbersome perceptual representations $\myPhi$, efficient methods are needed to address the computational and memory overload during training.



\subsection{Perceptually motivated time--frequency representations}

Spectrotemporal modulations (STM) are multiresolution, oriented oscillation rates in the time--frequency domain.
As one of the evidences that identify STMs as the underpinning for auditory perception, it is shown in neurophysiological research that systematic degradation of STMs in complex signals such as speech and music,  are correlated with a loss of intelligibility \cite{albouy2020}. 
Relatedly, spectrotemporal receptive fields (STRF) reflect each mammal's selectivities to STMs of differing scales (amplitude modulation) and rates (frequency modulation).
They are measurable through exposing the mammals' primary auditory cortex (A1) under systematically varied, parametrized auditory stimuli, such as ripples \cite{depireuxripples}. 
Based on these experimental results, \cite{Chi2005} among others have proposed computational idealizations of the A1 processing aiming to simulate STRF. 
These models translate acoustic signals to basic aspects of physiological responses, paving way towards constructing a computational proxy for indicating sonic differences. More recently, \cite{thoret2021} applied metric learning onto STM model responses and demonstrated its correlation to similarity judgements of musical sounds. 

Despite STRF's physical relevance, its numerical representation is cumbersome and RAM-intensive. It is also challenging to interface with common deep learning framework as a learning objective, due to a lack of differentiable implementation. One alternative approach is the joint time--frequency scattering transform (JTFS) proposed in \cite{andenjtfs2019}. JTFS is an operator that analyzes spectrotemporal modulation via joint wavelet convolutions, nonlinearities, and low-pass filters.  
Despite being close to STRF, its design results in more lightweight and efficient perceptual encodings without compromising the comprehensiveness of the captured information. 
JTFS succeeds as audio feature in various state-of-the-art classification tasks thanks to its time-shift invariance property, and is able to recover complex audio signals via gradient descent \cite{andenjtfs2019}, an indication of its high efficiency and preservation of information resolution. 
Meanwhile, in the spirit of \cite{thoret2021}, \cite{lostanlen2021time} reported strong correlation between the Euclidean distance in JTFS coefficients, and human subjects' similarity judgements on musical instruments' playing techniques.
Recently,  \cite{muradeli2022differentiable} published a differentiable implementation of JTFS, making it possible to be adopted as a learning objective.

\subsection{P-loss and Differentiable Spectral Loss}
Given a DNN $\myfW$ to retrieve $\tilde{\mytheta}$ from $\myx$, choosing the training objective largely influences learning dynamics and in turn the converged solution.

When neither $\myPhi$ nor $\myg$ is differentiable, a common approach termed parameter loss, or ``P-loss" for short, optimizes $\Vert \mytheta-\tilde{\mytheta} \Vert^2$ the mean squared error between the ground truth and estimated parameters. 
P-loss is a fast alternative as it circumvents the backward pass through $\myPhi$ and $\myg$. 
Yet, it only finds perceptually acceptable solution under the premise that $\Vert \tilde{\mytheta}-\mytheta \Vert \propto \Vert (\myPhi \circ \myg) (\tilde{\mytheta})-(\myPhi\circ \myg)(\mytheta)\Vert$; that is, a proximity in parameter space is equivalent to a proximity in the perceptual space. 
Indeed, parameter and perceptual loss surface share a common global minimum $\boldsymbol{\myW}^*$ such that $\boldsymbol{f}_{\boldsymbol{\myW}^*}({\myx})=\mytheta$.
However, the smoothness of the error surfaces should also align to converge to similar local minima \cite{yeeking2018}. 
In practice, a certain operator $\boldsymbol{h}$ is often applied onto $\mytheta$ to better approximate the relative distance in a perceptual dimension. Common operations include min--max scaling, logarithmic transformation, and other scaling methods based on known ranges or distributions of the parameter dimensions. The goal is to have the error in reparameterized space vary consistently as the perceptual discrepancy. 
These domain specific knowledge should also condition the sampling of datasets to ensure the success of a data-driven approach \cite{yeeking2011}. 
 
The benefit of P-loss is not limited to its role  as an inexpensive surrogate to perceptual loss. Being a convex objective in the parameter space, convergence of $\myW$ is easily achieved when parameters are normalized into similar ranges. On the contrary, perceptual loss does not guarantee this convergence. 
A typical spectral representation used in DDSP is the multiscale spectrogram distance (MSS). Given a target sound $\myx$ and estimated sound $\tilde{\myx}=(\myg \circ \myfW)(\myx)$, MSS\footnote{The original definition of MSS exist combined an $L^1$ norm between magnitudes and an $L^1$ norm between log-magnitudes. However, we choose to retain the $L^2$ norm instead, for consistency with Equation \ref{eq:psm} and \ref{eq-ddsp}.} is defined by: 
\begin{equation}
    \mathcal{L}_{\myx}^{\mathrm{MSS}}(\myW)=\sum_{k=5}^{10} \Vert \myPhi_{\mathrm{STFT}}^k(\myx)-\myPhi_{\mathrm{STFT}}^k(\tilde{\myx}) \Vert_2^2,
\end{equation}
where $\myPhi_{\mathrm{STFT}}^i$ refers to short-time Fourier transforms computed with window sizes of $2^k$.
Using $\mathcal{L}_{\myx}^{\mathrm{MSS}}$ as spectral loss is computationally inexpensive, does not require the presence of ground truth $\mytheta$, and more importantly reflects distance between harmonic and stationary sounds of the same pitch.
However, its adequacy is compromised when comparing sounds with different pitches \cite{turian2020}, inharmonicity \cite{han2023pnp} or nonstationarities \cite{vahidi2023meso}. 
When representing these sounds, MSS no longer accurately distinguishes large perceptual differences, and therefore fails to provide informative gradient vectors.

Due to the complementary nature between the above two approaches, choosing the right learning objective requires careful consideration on the characteristics of sounds and preprocessing of parameter space in each specific application. 
Encouragingly, attempts at merging the above two approaches have shown to be successful \cite{masuda2023}.
Nevertheless, these approaches usually require separate pre-training and fine-tuning stages, introducing auxiliary hyperparameters in the training scheme.




\section{Methods}
\subsection{Taylor expansion}
Given a spectral representation $\myPhi$ and a synthesizer $\myg$, a DDSP-style $L^2$ spectral loss can be expressed as 
\begin{align}
    \mathcal{L}^{\mathrm{DDSP}}_{\mytheta}(\mathbf{W}) &=
   \Vert \myPhi(\myx) - \myPhi\circ\myg \circ \myfW (\myx) \Vert^2_2
    \nonumber \\
    &=
   \big\Vert(\myPhi \circ \myg)(\mytheta)
    - (\myPhi \circ \myg)(\tilde{\mytheta})
    \big\Vert^2_2
    \label{eq-ddsp}
\end{align}

By relying on the assumption that the function $(\myPhi\circ \myg)$ is continuously differentiable, we perform its second-order Taylor expansion around $\tilde{\mytheta}$ and obtain
\begin{align}
    (\myPhi \circ \myg) (\tilde{\mytheta})  &= 
    (\myPhi \circ \myg) (\mytheta) +
    \mynabla_{(\myPhi\circ\myg)}(\mytheta) \cdot
    (\tilde{\mytheta} - \mytheta) \nonumber \\ &+
    O(\Vert \tilde{\mytheta} - \mytheta \Vert^2_2),
    \label{eq:phi-xtilde}
\end{align}
where the Jacobian matrix $\mynabla_{(\myPhi\circ\myg)}(\mytheta)$ has $P=\dim(\myPhi\circ\myg)(\mytheta)$ columns and $J=\dim\mytheta$ rows. 
From Equation \eqref{eq:phi-xtilde}, we deduce\footnote{The bracket notation $\big\langle \boldsymbol{y} \vert \mathbf{A} \vert \boldsymbol{y} \big\rangle$ is equivalent to $\boldsymbol{y}^{\top}\mathbf{A}\boldsymbol{y}$.}:
\begin{align}\mathcal{L}_{\mytheta}^{\mathrm{DDSP}}(\mathbf{W}) 
    &=
    \big\langle
    \tilde{\mytheta} - \mytheta
    \big\vert
    \myM(\mytheta)
    \big\vert
    \tilde{\mytheta} - \mytheta
    \big\rangle
    + O\big(\Vert \tilde{\mytheta} - \mytheta \Vert^3_2\big)
\end{align}
where, for every $\theta \in \Theta$,
\begin{equation}
    \myM(\mytheta) =\mynabla_{(\myPhi\circ\myg)}^{\top}(\mytheta)
    \mynabla_{(\myPhi\circ\myg)}(\mytheta)
    \label{eq:riemannian-metric}
\end{equation}
is a positive semidefinite matrix with $J$ rows and $J$ columns.
The optimal quadratic approximation of $\mathcal{L}_{\mytheta}^{\mathrm{DDSP}}$ near $\myW$ is
\begin{equation}
\mathcal{L}_{\mytheta}^{\mathrm{PNP}}(\myW)=\big\langle
    \tilde{\mytheta} - \mytheta
    \big\vert
    \myM(\mytheta)
    \big\vert
    \tilde{\mytheta} - \mytheta
    \big\rangle
    \label{eq:PNP}
\end{equation}

The main interest in PNP resides in its computational efficiency.
Since $\myPhi$ is  computationally expensive, gradient backpropagation through $\mathcal{L}_{\mytheta}^{\mathrm{DDSP}}$ is impractically slow. According to the chain rule, optimizing against conventional spectral loss requires evaluating at every iteration the gradient of each path in the spectral representation $\myS_p$ with respect to each neural net weight scalar $\mathbf{W_i}$. 
Using PNP, however, we decompose the above gradient computation into the matrix multiplication of two Jacobian matrices of $P \times J$ and $J \times |W|$ entries.
\begin{equation}
\mathbf{\mynabla}_{\myPhi \circ \myg \circ \mathbf{f}\circ \myg_{\mytheta}}(\mathbf{W}) =
    \mynabla_{(\myPhi\circ \myg)}(\tilde{\mytheta})\mynabla_{\mathbf{f}\circ \myg_{\mytheta}}(\mathbf{W})
\end{equation}
We precompute only once $\mynabla_{(\myPhi\circ \myg)}(\mytheta)$. 
Note that in practice, $J^2 \ll JP$: spectral representation lies in much higher dimension than synthesis parameters.
Therefore, despite the high dimensionality of $\mynabla_{(\myPhi \circ \myg)} (\mytheta)$, its inner product $\myM(\mytheta)$ is lightweight to be stored in memory and loaded on the fly while training.

\subsection{Riemannian geometry}
\label{secriman}
$\myM(\mytheta)$ defines an inner product on the tangent space of $(\myPhi\circ\myg)$ at each point $\mytheta$.
Alternatively viewed, it is the Riemannian metric induced by the smooth manifold composing differentiable map $(\myPhi\circ\myg)$ and the open set $\Theta \subset \mathbb{R}^{J}$.
Such a metric allows for measuring properties such as curvature, volume, and length in the manifold space.
Using Riemannian metric to investigate and exploit information geometry in a manifold space is not new. 
In this context, a relevant prior publication in \cite{chadebec2023}, used the inverse determinant of Riemannian metric to infer data importance and inform geometry-aware data sampling.

Let $\sigma_j^2(\mytheta)$, $v_j(\mytheta)$ be the paired eigenvalues and orthonormal eigenvectors of $\myM(\mytheta)$.
For each $v_j$, one has: $\myM(\mytheta) v_j(\mytheta) =\sigma_j(\mytheta)^2 v_j(\mytheta)$.
Hence, Equation \eqref{eq:PNP}
may be rewritten as 

\begin{equation}
    \mathcal{L}_{\mytheta}^{\mathrm{PNP}}(\myW) = 
    \sum_{j=1}^{J} \sigma_j^2(\mytheta)
    \big\vert
    \langle
    \tilde{\mytheta} - \mytheta
    \big\vert
    \boldsymbol{v}_j(\mytheta)
    \rangle
    \big\vert^2
    \label{eq:pnpevec}
\end{equation}

Indeed, if $\myM(\mytheta)$ is an identity matrix, $\sigma_j(\mytheta)^2=1$, $v_j(\mytheta)$ is the standard basis, and Equation \eqref{eq:pnpevec} falls back to P-loss. 
Yet, a Riemmanian metric $\myM(\mytheta)$ projects parameter error vectors onto $v_j(\mytheta)$, scale by $\sigma_j(\mytheta)^2$, and recombine them to form a loss objective which takes into consideration the non-Euclideanness of the multidimensional perceptual space.



\subsection{Levenberg-Marquardt algorithm}
\label{seclma}
In practice, $\myM(\mytheta)$ is not necessarily positive definite. Under cases where $\mytheta$ has interdependent dimensions, or when certain dimensions of the manifold space $(\myPhi \circ \myg)$ are invariant to changes in $\mytheta$, $\myM(\mytheta)$ is rank deficient. 
Such a condition in the Riemmanian metric poses a serious problem, as implied in the aforementioned metrics of inverse determinant: the determinant would collapse to zero, the metrics would blow up to infinity, implying an explosion of information containment.
Consequently even though $\mathcal{L}_{\mytheta}^{\mathrm{PNP}}$ remains strictly non-negative, designating it as a loss function faces a challenge.
Since $\myM(\mytheta)$ is positive semidefinite, a Cholesky decomposition leads to $\myM(\mytheta)=\mathbf{A}(\mytheta)^{\top}\mathbf{A}(\mytheta)$.
It follows that minimizing Eq. \ref{eq:PNP} is equivalent to minimizing $\Vert \mathbf{A}(\mytheta)(\tilde{\mytheta}-\mytheta)
\Vert_2^2$.
The corresponding ordinary least squares (OLS) estimator via Gauss-Newton Method is
\begin{align}
    \tilde{\mytheta}&=(\mathbf{A}(\mytheta)^{\top}\mathbf{A}(\mytheta))^{-1}\mathbf{A}(\mytheta)^{\top}\mathbf{A}(\mytheta)\mytheta \nonumber \\
    &=\myM^{-1}(\mytheta)\mathbf{A}(\mytheta)^{\top}\mathbf{A}(\mytheta)\mytheta.
\end{align}
Here, the inversion of $\myM(\mytheta)$ is ill-conditioned when $\myM(\mytheta)$ has a high condition number 
\begin{equation}
\kappa(\mytheta) = \frac{\max_{1\leq j\leq J}{\sigma_j(\mytheta)}}{\min_{1\leq j\leq J}{\sigma_j(\mytheta)}}.
\end{equation}
More precisely, a rank deficient matrix has at least one eigenvalue of zero, rendering its condition number to infinity and its inverse undefined. 
In the Levenberg-Marquardt algorithm, a regularization term is added to the matrix before inversion to avoid this potential instability.
\begin{equation}
    \tilde{\mytheta}=(\mathbf{A}(\mytheta)^{\top}\mathbf{A}(\mytheta) + \lambda \mathbf{I})^{-1}\mathbf{A}(\mytheta)^{\top}\mathbf{A}(\mytheta)\mytheta
\end{equation}
Each eigenvalue of $\myM$ increases by
the corresponding eigenvalue of the added diagonal matrix, in this case 
a constant factor $\lambda$. Consequently, the condition number $\kappa$ decreases and a numerically stable inverse can be found.
We take inspiration from LMA and adopt a similar corrective term approach to improve PNP. We define: 
\begin{align}
    \mathcal{L}_{\mytheta,\lambda}^{\mathrm{PNP}}(\myW)
    &=\big\langle
    \tilde{\mytheta} - \mytheta
    \big\vert
    \myM(\mytheta) + \lambda \mathbf{I}
    \big\vert
    \tilde{\mytheta} - \mytheta
    \big\rangle \nonumber \\ 
    &= \big\langle 
    \tilde{\mytheta} - \mytheta
    \big\vert
    \myM(\mytheta)
    \big\vert
    \tilde{\mytheta} - \mytheta
    \big\rangle + \lambda \Vert \tilde{\mytheta}-\mytheta\Vert^2
    \label{eq:PNP_corrected}
\end{align}

The corrective term simultaneously shifts the eigenvalues $\sigma_j^2$ by $\lambda$ thereby improving the condition number of $\myM$, and acts as a $L^2$ regularization term for the inverse problem.
In the limit of $\lambda \to \infty$, $\lambda$ balances the magnitudes of $\sigma_j^2$ such that \eqref{eq:PNP_corrected} approximates a P-loss regime. 

Similar to LMA, we initialize $\lambda$ at a large value $\lambda_{\mathrm{max}}=\mathrm{max}_{1\leq j\leq J}\sigma_j^2(\mytheta)$, the largest eigenvalue of all $\myM(\mytheta)$. 
We iteratively reduce it whenever epoch validation loss decreases, until reaching a low threshold (typically zero). This way, the optimization starts in the P-loss regime and gradually transitions to the perceptual loss regime. 
Another reason why this damping is beneficial is that the higher order error term in \eqref{eq:phi-xtilde} ceases to be negligible when $(\tilde{\mytheta}- \mytheta)$ is large. As the DNN weights are randomly initialized, $\tilde{\mytheta}$ is not guaranteed to approximate $\mytheta$ at the beginning of training. 
The network initialization is therefore crucial to the reliability of such a spectral loss approximation.



\section{Application}

\subsection{Joint time--frequency scattering transform ($\myPhi$)}

Let $\psi(t)$ be the mother wavelet, a complex-valued zero-average bandpass filter.  
Through dilating and scaling $\psi$ in time by a factor of $2^{\lambda}$, we may construct a filterbank of multiscale wavelets with a constant quality factor, i.e. constant ratio between their center frequencies and bandwidths
\begin{equation}
    \psi_{\lambda}(t) = 2^{\lambda}\psi(2^{\lambda}t),
\end{equation} where $\lambda\in \mathbb{R}$ are the log-frequency variables of the wavelet filterbank.
Its center frequencies are logarithmically spaced in the hearing range with $Q$ number of filters per octave, and $J$ number of octaves. Here, we set $Q=12$. 
Given a signal $x$, we first obtain the modulus scalogram after complex convolution with $\psi_{\lambda}(t)$, then we apply a low-pass filter $\phi_T(t)$ of time support $T$ which introduces invariance to time shifts within $\tau\leq T$. This results in the first-order scattering coefficients $\myS_1\myx$.
\begin{equation}
    \myS_1\myx = |\myx*\psi_{\lambda}(t)| *\phi_T(t)
\end{equation}
To recover the high frequency information lost in the time-averaging operation and further capture spectrotemporal modulations in the signal, we apply a second wavelet transform. 
We define two separate filterbanks $\psi_{\mu}(t)$ and $\psi_{l}(\lambda)$ in the time and log- frequency domain respectively. 
Their center frequencies refer roughly to the modulation scale and rate in STRF \cite{Chi2005}.
We adopt the 2D wavelets resulting from the outer product between them:
\begin{equation}
    \psi_{\mu,l,s} = \psi^{(t)}(2^{\mu}t) \psi^{(f)}(s2^l\lambda)
\end{equation}
, where $s=\pm 1$ is the ``spin'' parameter indicating the upward or downward directionality of the wavelet. 
We take $Q=1$ for the second order wavelet filterbanks, as the modulation frequency decomposition does not require the same frequency resolution.
The second order coefficients are obtained in the same way after modulus nonlinearities and timeaveraging with a spectrotemporal low pass filter. 
\begin{equation}
   \myS_2\myx = | \myS_1\myx * \psi_{\mu,l,s}(t, \lambda) | *\phi_{T,F}(t,\lambda)
\end{equation}
The convolution along frequency axis makes it robust to preserve frequency-dependent temporal structures and time--frequency warping even after the compulsory low pass filter for time-shift invariance. 
Finally, JTFS concatenates $\myS_1\myx$ and $\myS_2\myx$ to form the final time-invariant representation $\mathrm{JTFS}(\myx)=[\myS_1\myx, \myS_2\myx]$.

Previously demonstrated to improve discriminability among signals with small scattering coefficients in classification tasks \cite{muradeli2022differentiable}, we further apply a stabilized logarithmic transformation:
\begin{equation}
    \Phi_{\mathrm{JTFS}}(\myx)=\log\left(1+\frac{ \mathrm{JTFS}(\myx)}{\varepsilon}\right),
\end{equation}
where we have set the hyperparameter $\varepsilon$ to $10^{-3}$.

\subsection{AM/FM arppegiator ($\myg$)}
\label{amchirpsec}
Nonstationary signals governed by regularities in the mesostructural level are challenging to be effectively distinguished by first-order time--frequency representation focusing on local variations \cite{vahidi2023meso}. 
The AM/FM arppegiator is an amplitude-modulated ascending chirp signal:
\begin{align}
    \myg(\mytheta) = \sin\left(\frac{2\pi f_c}{\gamma \log 2} (2^{\gamma t}-1)\right) \sin(2\pi f_m t)\boldsymbol{\phi}(\gamma t),
\end{align}
where $\mytheta=\{\log f_c, \log f_m, \log \gamma \}$.

Modified by chirp rate $\gamma$ in octaves per second, a carrier signal sweeps exponentially over the frequency interval $[f_c, f_c2^{\gamma T}]$ in a period of time $T$. 
Then, a slow oscillation sinusoid of frequency $f_m$ in Hz is multiplied to modulate the amplitude envelope. 
Finally, we apply a Gaussian window of width
$0.2/\gamma$ in seconds onto the signal.

To construct the dataset, we sample each parameter from their respective intervals: $f_c\in [512,1024]$ Hz, $f_m\in [4,16]$ Hz and $\gamma \in [0.5,4]$ Hz, and generate sounds at a sampling rate of $2^{13}$ Hz. Specifically, we divide each dimension of the 3-D cube formed by these intervals into 30 equal steps in the logarithmic space, making $30^3=27\cdot 10^3$ three-dimensional grids.
Each audio sample is thus randomly sampled from a point in each grid. 
To ensure varied distribution in the training and validation set, we take the middle $14^3\approx 2.7\cdot 10^3$ grids as the validation set. The training and testing sets remain in the same distribution at the peripheral of the sampling cube. The train/test/val split is $8:1:1$. 

\subsection{Drum sound synthesizer ($\myg$)}

Another type of nonstationary signals can be exemplified by inharmonic transient sounds, such as those coming from a percussive instrument. 
We synthesize percussive sounds via physical model of a 2D membrane residing in the Cartesian coordinate system $\myu$, where a random position's vertical displacement $\mathbf{X}(t,\myu)$ is governed by a fourth-order damped partial differential equation. For $t\geq 0$, 
\begin{align}   
    \left(\dfrac{\partial^2 \mathbf{X}}{\partial t^2}(t,\myu)
    - c^2 \nabla^2\mathbf{X}(t,\myu) \right)
    &+ S^4 \big(\nabla^4
    \mathbf{X}(t,\myu)\big)
    \nonumber \\
    +\dfrac{\partial}{\partial t}
    \Big(d_1 \mathbf{X}(t,\myu) + d_3 &\nabla^2\mathbf{X}(t,\myu)\Big) =
    0
    \label{eq:pde}
\end{align}

The higher order spatial derivatives and first-order time derivative account for materiality, air-induced frictions and energy dissipation respectively, introducing in the resulting solution inharmonicities and uneven decay envelopes among partials. 
Furthermore, the characteristics of such a solution are parameterized by material stiffness $S$, traveling wave speed $c$, frequency-independent damping coefficient $d_1$ and frequency-dependent damping coefficient $d_3$.
For compatibility with the adopted coordinate system, we focus on a rectangular drum of side lengths $l$ by $l\alpha$ with its boundaries at zero at all time. 
We simulate the excitation by setting the initial condition at $t_0=0$ to be $\mathbf{X}(t_0,\myu=(0.4l, 0.4l\alpha)) = 0.03 \text{ meters}, \text{and } 0 \text{ otherwise}$.
 
We adopt the functional transformation method (FTM) \cite{trautmannbookftm2003}, a solver to the above PDE, as our synthesizer $\myg(\mytheta)$, where $\mytheta = \{S, c, d_1, d_3, \alpha\}$. FTM disentangles time-space solution by performing Laplace and Sturm-Liouville transforms and thereby transforming the variables into their respective functional spaces. There, algebraic solutions in the functional space domain are obtained and then inverse transformed back into the time-space domain.
The solution follows a modal synthesis form:
\begin{align}
    \myx(t) = \mathbf{X}(t,\myu)
    = 
    \sum_{\bold{m}\in\mathbb{N}^2}
    K_m(\myu,t) \exp(\sigma_\bold{m} t) \sin(\omega_{\bold{m}} t).
\end{align}
The coefficients $K_\bold{m}(\myu,t)$, $\sigma_\bold{m}$, $\omega_\bold{m}$ are respectively modal amplitudes, modal exponential decay rates, and modal frequencies, where $\bold{m}=(m_1,m_2)$. They are derived from the original PDE parameters in the following ways.
\begin{equation}
    \omega_{\bold{m}}^2 = \left(S^4-\frac{d_3^2}{4}\right)\Gamma_{\bold{m}}^2 +
    \left(c^2+\frac{d_1d_3}{2}\right)\Gamma_{\bold{m}}
    - \frac{d_1^2}{4}
\end{equation}
\begin{equation}
     \sigma_{\bold{m}} = \frac{d_3}{2}\Gamma_{\bold{m}}-\frac{d_1}{2}
\end{equation}
\begin{equation}
     K_{\bold{m}}(\myu,t) = y_{u}^{m}\delta(t)\sin\left(\frac{\pi m_1 u_1}{l}\right)\sin\left(\frac{\pi m_2 u_2}{l\alpha}\right)
\end{equation}
where $\Gamma_{\bold{m}}=\pi^2m_1^2/l^2+\pi^2 m_2^2/(l\alpha)^2 $, and $y_{u}^{\bold{m}}$ is the ${\bold{m}}^{th}$ coefficient associated to the eigenfunction $\sin(\pi m \myu/l)$ that decomposes $y_u(\myu)$.
Lastly, to relate better input parameters to sonic characteristics, we reparameterize $\mytheta = \{S, c, d_1, d_3, \alpha\}$ to $\mytheta = \{\log\omega_1, \tau_1, \log p, \log D, \alpha\}$ in the following way:
\begin{align}
    \omega_1 & = \sqrt{\frac{\beta^4}{\alpha^2}S^4+\frac{\beta}{\alpha}c^2-\frac{1}{4}(\frac{\beta}{\alpha}d_3-d_1)^2} \\
    \tau_1 & = \frac{2}{d_1-\frac{\beta}{\alpha}d_3}\\
    p & = \frac{d_3}{\beta d_3 - \alpha d_1}\\
    D & = \frac{1}{\alpha \omega_1}\sqrt{S^4-\frac{d_3^2}{4}}
\end{align}

where $\beta = 1 + 1/\alpha$.
For simplicity, we denote the first mode in both dimension $m_1=m_2=1$ as $\bold{m}=1$.
Fundamental frequency $\omega_1$ ranges between $40$ Hz and $1$kHz; duration $\tau_1$, between $400$ ms and $3$ s; inhomogeneous damping rate $p$, between $10^{-5}$ and $0.2$; frequential dispersion $D$, between $10^{-5}$ and $0.3$; and aspect ratio $\alpha$, between $10^{-5}$ and $1$.
These predefined ranges motivated by perception can be converted back to the PDE parameters in the following way:
\begin{align}
    S & = \left(\left(D\omega_1\alpha\right)^2+\left(\frac{p\alpha}{\tau_1}\right)^2 \right) ^{\frac{1}{4}} \\
    c & = \sqrt{\frac{\alpha}{\tau_1^2}\left(\frac{1}{\beta}-p^2\beta\right)+\alpha\omega_1^2\left(\frac{1}{\beta}-D^2\beta\right)}\\
    d_1 & = \frac{2}{\tau_1}(1-p\beta)\\
    d_3 & = -\frac{2p\alpha}{\tau_1}
\end{align}
The dataset is constructed the same way as described in section \ref{amchirpsec}. We take $m_1=m_2=10$ and synthesize $\sim3$s sounds at a sampling rate of 22050 Hz. The hyperdimensional cube is divided into 10 equal steps per parameter dimension, forming 100k samples in total. The train/test/validation split is $8:1:1$. 

\subsection{Reparametrization ($\mytheta$)}
\label{secparamscale}
To retrieve $\mytheta$, the decoder $\myfW$ commonly ends with a linear layer of the same size activated by certain function. 
Scaling of ground truth parameters is a key factor in designating such an activation function and in turn the ease of optimization. 
In this work, we examine the effect of two modes of reparametrization on the neural net's learning dynamics - min--max scaling and logarithmic transformation. 

\begin{table}[]
    \centering
    \begin{tabular}{c|c|c}
    \textbf{Min-Max} & \textbf{Log Transform} & \textbf{Activation} \\
    \hline    
   $\checkmark$ & $\checkmark$ & Tanh \\
    \hline
    $\checkmark$ & $\times$ & Tanh \\
    \hline
    $\times$ & $\checkmark$ & None \\
    \hline
    $\times$ & $\times$ & LeakyReLU \\
    \end{tabular}
    \caption{Activation functions used by the last linear layer of $\myfW$ under different reparametrizations}
    \label{tab:paramscale}
\end{table}

Min--max scaling is a common technique to normalize parameters into a confined range: typically, $[0,1]$ or, as in our case, $[-1, 1]$. 
\begin{equation}
    h_{\mathrm{minmax}}(\mytheta) = -1 + 2\frac{\mytheta - \mytheta_{\mathrm{min}}}{\mytheta_{\mathrm{max}}-\mytheta_{\mathrm{min}}}
\end{equation}
Logarithmic transformation $h_{\mathrm{log}}(\mytheta) = \log(\mytheta)$
 applies log operations onto parameters that influence perception in an approximately logarithmic scale and or have too small ranges.
In our case, these parameters are $f_c$, $f_m$, and $\gamma$ for the AM/FM arpeggiator; and  $\omega_1$, $p$, and $D$ for the drum synthesizer. 

Both scaling operations above are considered as domain knowledge that modify the optimization landscape such that a more stable convergence is achieved. To finetune an output vector of entries lying or affecting the loss objective in drastically different ranges, the Jacobian matrix $\mynabla_{(\myPhi \circ \myg)}(\tilde{\mytheta})$ needs to produce gradient updates with magnitudes of the according sensitivities. This is hard to achieve as, in deep learning, the commonly adopted first-order optimizers do not consider curvature of the optimization landscape.
This poses a risk of perpetually improving one parameter's accuracy at the expense of detrimenting the other. 
Furthermore, optimizing at once parameters of different ranges is problematic for unweighted loss functions. 
Consider optimizing $\myfW$ against P-loss, an error in the parameter with large range contributes more significantly to the overall loss, while those with small ranges have minimal effect.
Consequently the network naturally tends to prioritize reducing errors in the largest parameter.

We apply none, one of or both scaling methods to the ground truth parameters, while adapting the final layer accordingly as specified in Table \ref{tab:paramscale}.


\subsection{Learned Decoder ($\myfW$)}
A suite of DNN architecture known as EfficientNet \cite{tan2019efficientnet}, uses compound scaling method to find the optimal balance for widths, depths and input resolutions of subsequent convolutional blocks.
The network achieved state-of-the-art classification result on both image \cite{tan2019efficientnet} and audio tasks \cite{muradeli2022differentiable} using significantly less parameters.
We adopt the most light weight version EfficientNet-B0 as our encoder $\myfW$ architecture, summing to 4M trainable parameters. 
The network takes in constant Q transform (CQT) coefficients of each audio sample, and outputs a vector of dimension $\dim{\mytheta}$. The final linear layer vary in its activation functions according to the chosen reparametrization, specified in Table \ref{tab:paramscale}.
We leave to future work the question of scaling our experiment benchmark beyond B0 and up to B7.

\section{Experiments}
\begin{table*}[ht]
    \centering
    \begin{tabular}{r|ll|lr|lr}
     \multicolumn{3}{c}{} & \multicolumn{2}{c}{\makecell{AM/FM arpeggiator}} & \multicolumn{2}{c}{\makecell{FTM drum synthesizer}} \\ 
     \hline
    \makecell{} & \makecell{Loss} & \makecell{\myPhi{}} & \makecell{JTFS distance} & \makecell{MSS} & \makecell{JTFS distance} & \makecell{MSS} \\
    
    \hline
     Han \emph{et al.} 2020 \cite{han2020wav2shape} & P-loss & ---  
    & \phantom{0}$5.2$ $\pm$ $0.3$ & $0.36$ $\pm$ $0.01$ & \phantom{0}$47$ $\pm$ $7$ & $0.97$ $\pm$ $0.07$\\
     Engel \emph{et al.} 2020 \cite{engel2020ddsp} & $\mathcal{L}_{\theta}^{\mathrm{DDSP}}$ & $\myPhi{}_{\mathrm{MSS}}$  &  $47\phantom{.0}$ $\pm$ $3\phantom{.0}$  & $1.1$ \phantom{.0}$\pm$ $0.09$ & \phantom{0}$87$ $\pm$ $6$ & $1.3$ $\pm$ $0.09$\\
    Masuda \emph{et al.} 2023 \cite{masuda2023} & P-loss and $\mathcal{L}_{\theta}^{\mathrm{DDSP}}$ & $\myPhi{}_{\mathrm{MSS}}$ &  
 $14\phantom{.0}$ $\pm$ $8\phantom{.0}$ & $0.51$ $\pm$ $0.07$ & $90$ $\pm$ $5$ & $1.3$\phantom{0} $\pm$ $0.06$\phantom{0} \\
   
    Han \emph{et al.} 2023 \cite{han2023pnp} & $\mathcal{L}_{\theta}^{\mathrm{PNP}}$ & $\myPhi{}_{\mathrm{JTFS}}$  & \phantom{0}$5.5$ $\pm$ $0.7$ & $0.37$
 $\pm$ $0.04$ & \phantom{0}$59$ $\pm$ $1$ & $1.09$ $\pm$ $0.01$\\
   Ours & P-loss and $\mathcal{L}_{\theta}^{\mathrm{PNP}}$ & $\myPhi{}_{\mathrm{JTFS}}$ & \phantom{0}$\boldsymbol{3.2}$ $\pm$ $\boldsymbol{2}\phantom{.0}$ & $\boldsymbol{0.26}$ $\pm$ $\boldsymbol{0.06}$ & \phantom{0}$\boldsymbol{32}$ $\pm$ $\boldsymbol{1}$ & $\boldsymbol{0.78}$ $\pm$ $\boldsymbol{0.03}$\\

    \end{tabular}
    \caption{Average JTFS and MSS distance on the test set. Five loss objectives
    are trained to retrieve parameters from two sound synthesizers: AM/FM arppegiator (left) and FTM drum synthesizer (right). 
    The top four models are attempted reproductions of published papers referenced at the leftmost column, with slight  variations in the training procedure: \cite{engel2020ddsp} and \cite{masuda2023} have used a weighted combination of MSS and log-MSS L1 distance whereas we use an MSS $L^2$ distance; \cite{han2020wav2shape} and \cite{han2023pnp} did not apply a ``reduce learning rate on plateau'' policy.}
    \label{tab:results}
\end{table*}

\subsection{Competing Approaches}
\label{sec:opt}
For each sound synthesizer, we train $\myfW$ against 3 loss objectives: P-loss as \cite{han2020wav2shape}, $\mathcal{L}_{\mytheta,\lambda}^{\mathrm{PNP}}$ based on $\myPhi_{\mathrm{JTFS}}$ with adaptive damping as \cite{han2023pnp}, and MSS as \cite{engel2020ddsp}. Each network is exposed to 25600 samples per epoch for a total of 70 epochs. For P-loss and PNP loss we train with a batch size of 256 samples. For MSS loss, we need to reduce the batch size to 64 samples due to the memory constraints of $\myPhi \circ \myg$.
To eliminate the effect of differing batch sizes on experimental results, we also trained our models using P-loss and PNP loss with a batch size of 64 samples. As these supplementary experiments did not indicate significant degradation nor narrowed the advantageous leap from the results of MSS loss, we choose not to include them in Table \ref{tab:results}.
In making our best efforts to replicate \cite{engel2020ddsp, masuda2023}, all above experiments are trained with Adam optimizer and a scheduled learning rate initialized at 0.001 that reduces by a factor of 0.1 whenever validation loss plateaus for three epochs.
Finally in the spirit of \cite{masuda2023}, we propose to pretrain with P-loss for 35 epochs, and finetune with $\mathcal{L}_{\mytheta,\lambda}^{\mathrm{PNP}}$ for 35 epochs, using weight decay and gradient clipping in addition to the above optimizer setup. The full algorithm is written explicitly in Algorithm \ref{alg:cap}. 
The ground truth parameters are pre-processed with both min--max scaling and logarithmic transformation, unless specified otherwise.

\subsection{Evaluation}
We adopt two metrics to evaluate our experiments: MSS and JTFS distance. 
Both being perceptual metrics, the former assesses local frequency component alignments at various time scales. 
The latter examines additional differences in frequency and temporal modulation rates. For each trained model, we report both evaluation metrics averaged over the entire test set in Table \ref{tab:results}. 

\begin{figure*}
    \centering 
    \includegraphics[width=\textwidth]{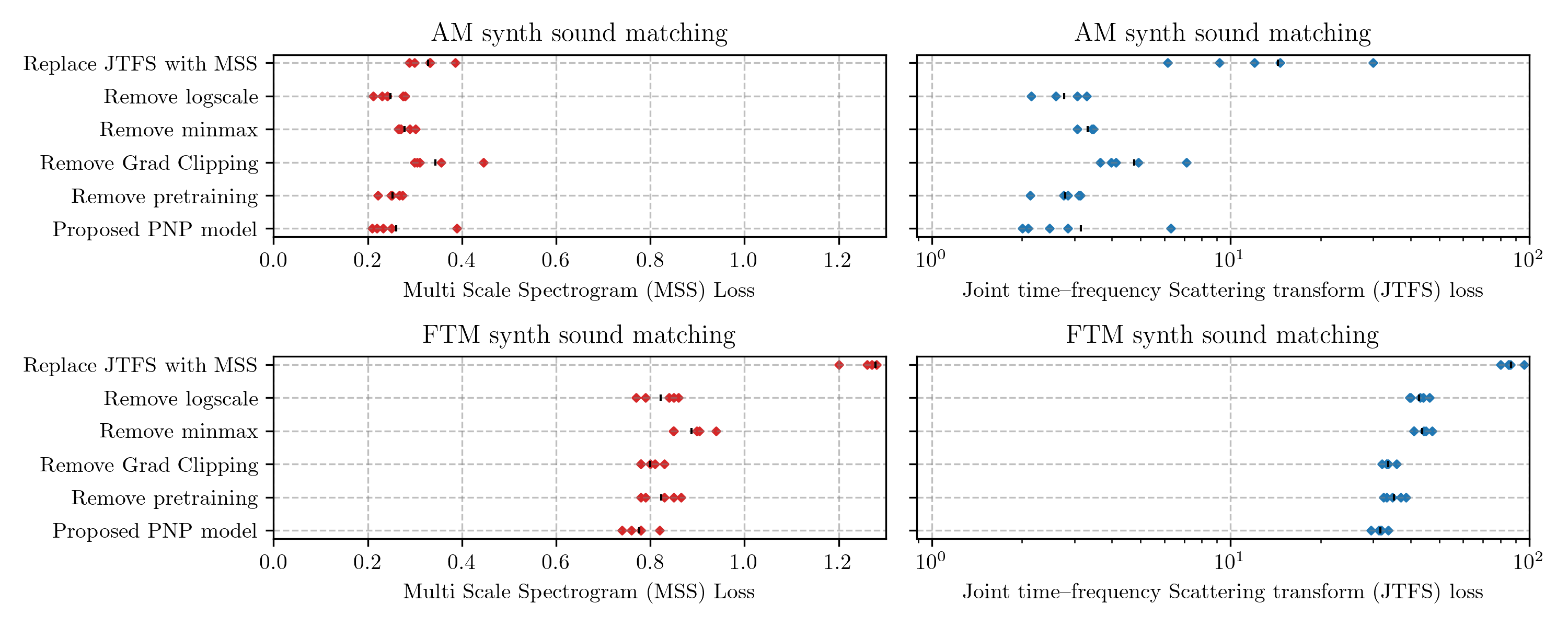}
   
    \caption{Ablation study of the proposed model. We demonstrate the deterioration of both JTFS metrics (left column) and MSS metrics (right column) under individual removal of one of various elements in the algorithm. For both metrics, lower values indicate a better perceptual accuracy.
    Bottom row is the proposed model detailed in Algorithm \ref{alg:cap}. Each row above shows the extent of degradation when disabling each mechanism.
    The use of JTFS as opposed to MSS is essential to model performance. Each model is randomly initialized with varying seeds and trained five times (red diamonds), with black vertical lines indicating the average metric across the trials.}
    \label{fig:ablation}
\end{figure*}

\subsection{Discussion}
Our proposed model consistently outperforms the other methods on both metrics.
P-loss and $\mathcal{L}_{\mytheta,\lambda}^{\mathrm{PNP}}$ also converge stably to suboptimal solutions. 
On the contrary, MSS fails to induce consistent gradient updates that retrieve accurate parameters, even under the help of improved model initialization.

\begin{algorithm}
\caption{Proposed algorithm for deep neural network training for perceptual sound matching with regularized JTFS-based PNP loss. Includes all additional refinements: P-loss pretraining, adaptive damping, gradient clipping, and reparametrization (both logarithmic and minmax).}\label{alg:cap}
\begin{algorithmic}
\State $\boldsymbol{Input}$: Neural network architecture $f$ with weights $W$, pairs of parameters and sounds $\{(\mytheta_n, \myx_n)\}_{n=1}^N$,
 minibatch size $B$, steps per epoch for training set $S_{\mathrm{tr}}$, and validation set $S_{\mathrm{val}}$, hyperparameters $\eta$, $\beta_1$, $\lambda$, $\varepsilon$, $\rho$. 
\State Set $\eta \gets 0.001$, $\beta_1 \gets 0.965$, $m_{T/2}\gets0$, $\epsilon \gets 10^{-15}$, $\rho\gets0.04$ 
,$W \sim$ Normal i.i.d.
\For{sample $n=1$ to $N$}
\State $\bar{\mytheta}_n \gets  h(\mytheta_n)$
\Comment{Section \ref{secparamscale}}
\State $\myM(\bar{\mytheta}_n) \gets \nabla_{(\myPhi \circ \myg \circ h^{-1})}^{\top}(\bar{\mytheta}_n)\nabla_{(\myPhi \circ \myg \circ h^{-1})}(\bar{\mytheta}_n)$
\EndFor 

\For{epoch $t=1$ to $T/2$}
\For{step $s=1$ to $S_{tr}$}
\State $\mathcal{L}_p \gets \frac{1}{BS_{\mathrm{tr}}}\sum\limits_{n\sim \bold{U}(1,N)} \Vert \bar{\mytheta}_n - \myfW(\myx_n)\Vert_2^2$ \Comment{P-loss}
\State $\myW \gets \myW - \frac{\eta}{BS_{\mathrm{tr}}}\sum\limits_{n\sim \bold{U}(1,N)} \nabla_{\myW}\mathcal{L}_p$
\EndFor 
\State $\mathcal{L}_p^{t} \gets \frac{1}{BS_{\mathrm{val}}}\sum\limits_{n\sim \bold{U}(1,N)} \Vert \bar{\mytheta}_n - \myfW(\myx_n)\Vert_2^2$
\If {$\mathcal{L}_p^t \geq \mathcal{L}_p^{t-3}$}
\State $\eta \gets 0.1 \eta$
\Comment{Learning rate scheduler}
\EndIf
\EndFor 

\State Set $\lambda \gets \lambda_{\max}(\myM(\mytheta_n))$
\Comment{Section \ref{seclma}}
\State Set $\eta \gets 0.001$
\For{epoch $t=T/2+1$ to $T$}
\For{step $s=1$ to $S_{tr}$}

\State $\delta_n \gets \myfW(\myx_n)-\bar{\mytheta}_n $
\State  $\mathcal{L}_{\mathrm{PNP}} \gets \frac{1}{BS_{\mathrm{tr}}}\sum\limits_{n\sim \bold{U}(1,N)}\langle \delta_n \vert \myM(\bar{\mytheta}_n)+\lambda\mathbf{I} \vert \delta_n \rangle$\\ \Comment{PNP}

\State $m_t \gets \beta_1 m_{t-1} + (1-\beta_1)\nabla_{\myW} \mathcal{L}_{\mathrm{PNP}}$ \Comment{Momentum}
\State $\myW \gets \myW-0.1\eta \myW$ 
\Comment{Weight decay}
\State $\myW \gets \myW -\eta \cdot \mathrm{clip} (m_t/\varepsilon,\rho)$
\Comment{Gradient Clipping}

\EndFor 
\State  $\mathcal{L}_{\mathrm{PNP}}^t \gets \frac{1}{BS_{\mathrm{val}}}\sum\limits_{n\sim \bold{U}(1,N)}\langle \delta_n \vert \myM(\bar{\mytheta}_n)+\lambda\mathbf{I} \vert \delta_n \rangle$
\If {$\mathcal{L}_{\mathrm{PNP}}^t \geq \mathcal{L}_{\mathrm{PNP}}^{t-3}$}
\State $\eta \gets 0.1 \eta$
\If {$\mathcal{L}_{\mathrm{PNP}}^t \leq \mathcal{L}_{\mathrm{PNP}}^{t-1}$}
\State $\lambda \gets 0.2 \lambda$
\Comment{Adaptive damping}
\EndIf
\EndIf

\EndFor 
\State 
\Return{$\myfW$}
\end{algorithmic}

\end{algorithm}

\subsubsection{Perceptual distance between nonstationary sounds}

The large discrepancy between using MSS versus JTFS-based loss objectives in retrieving parameters from nonstationary sounds is expected. It is also the pivotal factor contributing to the success of accurate parameter retrieval, as shown in the ablation study of Figure \ref{fig:ablation}. 

First, prior work such as \cite{turian2020, hayes2023sinusoidal} have shown that spectrogram distances are inaccurate in representing distance between sinusoids of different frequencies. This knowledge manifests in the difficulties encountered during retrieving $f_c$ from AM chirp signals and $\log\omega_1$ from FTM percussive sounds. In particular, \cite{han2023pnp} has shown how supplying ground truth frequency aids the retrieval of other parameters.

Second, AM chirp signals are governed by long-range periodicity in time ($f_m$) otherwise referred to as the "meso-structures", which are imperceptible by first order time--frequency analysis with short time support. \cite{vahidi2023meso} has visualized the loss surfaces and gradient fields to retrieve ${f_m,\gamma}$ from the same AM chirp sounds with known $f_c$ using MSS and JTFS loss objectives. It is observed that while MSS is able to provide decently reliable gradient updates on $\gamma$, its updates on $f_m$ are uninformative. 
On the contrary, the second order coefficients of JTFS analyze precisely the temporal modulation rate and thus provide informative distance measures that indicate errors in $f_m$. 

Third, in terms of frequency evolution, AM chirp signals have ascending single-tone frequency with their slopes dictated by $\gamma$. Drum sounds have $m_1m_2=100$ partials spaced out in exact or progressive deviation from the harmonic series, determined by dispersion $D$. While neither of these characteristics correspond exactly to a frequency modulation, single-tone frequency evolution is sufficient to be captured by first order time--frequency analysis and the partials spacings indeed have underlying frequency periodicity. That being said, inharmonic partial spacings pose challenge in modulation rate extraction for JTFS. 

Finally, although all synthetic sounds in this dataset are aligned in time, real-world sounds are rarely captured with such precision.
While JTFS is already a much more adequate objective in representing nonstationary sounds governed by spectrotemporal modulations, its potential can be further exploited with its time-shift invariance property, for example when applied onto more realistic sounds that misalign in time.

\subsubsection{Learning dynamics}

\begin{figure}
    \centering \includegraphics[width=\linewidth]{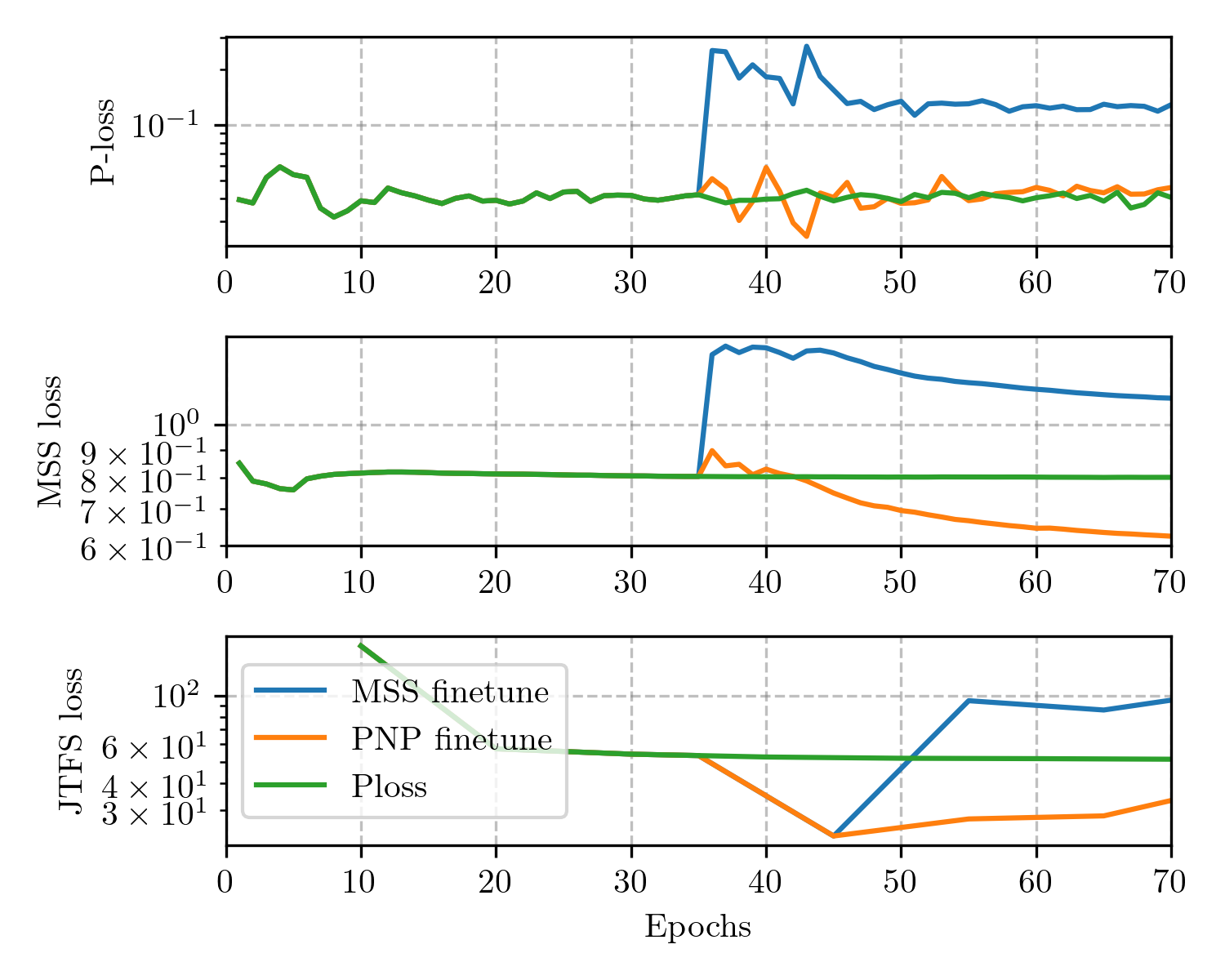}
    \caption{Learning dynamics of P-loss. Pretraining with P-loss followed by finetuning with JTFS-based PNP loss and MSS loss. The curves show progression of three measures (average P-loss, MSS, and JTFS distance on validation set) over the course of 70 epochs training time. 
    P-loss converges rapidly yet reaches plateau on all metrics. JTFS-based PNP finetuning keeps improving upon the two perceptual metrics without deteriorating on P-loss metrics. On the contrary, MSS finetuning fails to converge to an optimal solution after a spike in metrics due to a re-initialized learning rate.\label{fig:learningdynamics}}
\end{figure}

To trace the convergence of each loss objective over time, we train a model with P-loss for 35 epochs, then use either $\mathcal{L}_{\mytheta, \lambda}^{\mathrm{PNP}}$ or $\mathcal{L}_{\mytheta}^{\mathrm{DDSP}}$ to finetune the model, or continue training with P-loss for another 35 epochs. 
The experiments' optimization setups correspond to respectively \cite{han2020wav2shape}, \cite{masuda2023} and our proposed approach with weight decay and gradient clipping. 
We observe in Figure \ref{fig:learningdynamics} that model trained with P-loss soon converges and reaches plateau in both MSS and P-loss metrics. 
At the start of the fine-tuning stage, we see an abrupt leap in MSS metrics, most likely due to a mismatched learning rate.
We expect that the ``reduce learning rate on plateau'' mechanism would automatically address this discrepancy. 
Indeed despite the initial leap, $\mathcal{L}_{\mytheta,\lambda}^{\mathrm{PNP}}$ is able to catch up with P-loss model in parameter error and continue to improve upon MSS metrics. 
Nevertheless, models trained with $\mathcal{L}_{\mytheta}^{\mathrm{DDSP}}$ deteriorate instantly and fail to achieve a good solution.

\subsubsection{Effect of reparametrization}

Both min--max and logarithmic transformation are domain-specific knowledge that supposedly aids linearizing parameter errors in the perceptual error domain. We hope, through the inclusion of reparametrization, to provide a continuous optimization landscape where the same error magnitudes at different parameter regions reflect similar perceptual errors. 
As expected we observe in Figure \ref{fig:ablation} that both scaling operations help ease the retrieval. 
For both AM/FM arppegiator and FTM drum synthesizer, min--max scaling contributes more to the improvement of model convergence than does logarithmic transformation.
Even though removing logarithmic transformation results in inadequate scaling that further deviates from linear correlation with perceptual difference, the min--max scaling still confines the parameter range and in turn the prediction error. 
On the contrary, removing min--max scaling while keeping logarithmic transformation induces drastic parameter ranges and poses serious challenge in training the penultimate layer of $\myfW$, even if the logarithmic transformation reduces that range discrepancy. 
This potentially explains the relatively larger degradation in removing min--max scaling when retrieving from FTM drum sounds.


\subsubsection{Effect of optimizer}

The addition of weight decay and gradient clipping has greatly improved the convergence across all models, reducing both metrics by up to 2 folds.
As shown in Algorithm \ref{alg:cap} at each iteration, $\myW$ is weight decayed, and then updated by an exponential moving average of the gradient clipped by a scalar $\rho$. 
Both mechanisms aim to circumvent the presence of unusually large gradient updates in the current minibatch that causes unstable convergence or oscillations around local minimas. 

\subsubsection{Blending parameter and perceptual loss}
Despite the slight improvement, pretraining with P-loss is not instrumental in our proposed method as shown in Figure \ref{fig:ablation}.
Indeed, as $\lambda$ is initialized to be a big number, the corrected $\myM$ has uniform eigenvalues and thus train the decoder in the P-loss regime. The evolution of subsequent reduction of $\lambda$ also resembles a gradual transition to optimizing with a perceptual objective. 
Common practice of blending multiple learning objectives require hyperparameter tuning. In our case this is done semi-automatically by singular value decomposition of $\myM$.
As $\lambda_0$ is draw from data, namely initialized with the largest eigenvalue of all $\myM(\mytheta)$ in the dataset, the only hyperparameter needed is the damping rate that designate the rate of transitioning.

Experiments demonstrated that, for $\myg$ generating nonstationary sounds, complex $\myPhi$ such as JTFS is indispensable to the success of PSM. 
Reparametrization played a more important role in FTM drum synth parameter retrieval, which has difficult parametrization with more diverse parameter ranges and varied perceptual scaling. Removing pretraining with P-loss has little to no effect on model performance.  
The degradations among JTFS and MSS metrics are in general consistent. However, due to the presence and absence of second-order time--frequency analysis, the optimal solutions in JTFS metrics are not always the best in MSS metrics. This tendency is especially pronounced in AM synth, possibly due to its construction's exact alignment with JTFS analysis.
    
\section{Conclusion}
We have proposed perceptual neural physical (PNP) loss, a novel learning objective for PSM balancing computational efficiency and perceptual relevance. PNP has achieved state-of-the art sound matching results on two differentiable synthesizers of nonstationary sounds. Important improvements on the results involve the addition of adaptive damping mechanism, the use of JTFS as perceptual feature, reparametrization, gradient clipping and weight decay.

PNP is the optimal quadratic approximation of  $L^2$ perceptual loss in the synthesis parameter space. 
It precomputes and stores perceptual effects in the form of kernels in a bilinear formulation, therefore allows working with otherwise cumbersome differentiable implementations of auditory features.
Its adaptively damped variation enables semi-automatic transition between parameter loss and perceptual loss regimes.

Applying PNP objective calls for two prerequisites: a synthetic training set for supervised learning, and a differentiable synthesizer and perceptual representation for automatic differentiation. 
With the upsurge of direct and adapted implementations of differentiable synthesizers, the latter requirement is increasingly easy to fulfill in practice.
In contrast, the former requirement remains a limitation that refrains PNP from solving PSM with real world data. 
To address this current lack of supervision, future work, such as domain adaptation, is necessary to transfer the knowledge acquired by neural networks from exposure to synthetic data to applications on real world data.
\section*{Acknowledgment}

This work was granted access to the HPC resources of IDRIS under the allocation 20XX-AD011014253 made by GENCI.
V. Lostanlen is supported by ANR project nIrVAna (ANR-23-CE37-0025-04).

\ifCLASSOPTIONcaptionsoff
  \newpage
\fi



\bibliography{main.bib}
\bibliographystyle{IEEEtran}
\end{document}